\documentclass[aps,prl,twocolumn,groupedaddress,floatfix]{revtex4}

\usepackage{graphicx}% Include figure files
\usepackage{subfigure}% Include subfigures
\usepackage{epsfig}
\usepackage{amsmath}
\usepackage{bbm}

\def\be{\begin{equation}}
\def\ee{\end{equation}}
\def\bea{\begin{eqnarray}}
\def\eea{\end{eqnarray}}

\begin{document}

\title{Time-dependent evolution of two coupled Luttinger liquids}

\author{E. Perfetto}
%\author{}

\affiliation{Instituto de Estructura de la Materia.
              Consejo Superior de Investigaciones Cient{\'\i}ficas.
              Serrano 123, 28006 Madrid. Spain.\\
              and\\
             Istituto Nazionale di Fisica Nucleare - Laboratori
        Nazionali di Frascati, Via E. Fermi 40, 00044 Frascati, Italy.}

\affiliation{}

\date{\today}

\begin{abstract}
We consider two  disconnected Luttinger liquids which are coupled at
$t=0$ through  chiral density-density interactions. Both for $t<0$
and $t \geq 0$ the system is exactly solvable by means of
bosonization and this allows to evaluate analytically the
time-dependence of  correlation functions. We find that in the
long-time limit  the critical exponent governing the one-particle
correlation function differs from the exponent dictated by the
equilibrium ground state of the coupled system. We also discuss how
this reflects on some physical quantities which are accessible to
real experiments.

\end{abstract}
%\keywords{}

\maketitle
%\subsection{}
%\subsubsection{}
%%%%%%%%%%%%%%%%%%%%%%%%%%%%%%%%%%%%%%%%%%%%%%%%%%%%%%%%%%%%%%%%%%%
%%%%%%%%%%%%%%%%%%%%%%%%%%%%%%%%%%%%%%%%%%%%%%%%%%%%%%%%%%%%%%%%%%%

Time-dependent quantum systems are attracting much the attention
during the last years. This is because the dynamical response of
nanoscopic devices is nowadays accessible in many experiments with
high relevance in practical applications such as quantum computing
and single-electron transport\cite{prog}. On one side we mention the
wide investigation of out-of-equilibrium phenomena in transport
experiments, where time-dependent transient currents are measured at
picosecond time-scales\cite{timeqd}. On the other side, recent
experiments in ultracold atoms confined in optical lattices have
shown that it is possible to time-tune the strength of the
interactions in both bosonic\cite{opt1} and fermionic\cite{opt2}
systems, by using the so called Feshbach resonance. In these cases
it arises the intriguing question of what is the steady state to
which the initial ground state relaxes to\cite{hcb}\cite{rigol}.

In the light of this  challenging physics, the theoretical
investigation of the time-dependent evolution of many-body
interacting systems deserves special attention. Recently, an exact
formulation of time-dependent transport in electronic
systems\cite{cini} was derived in the framework of
time-dependent-density-functional-theory\cite{stef}. Unfortunately
its implementation to strongly interacting systems results hard.
An efficient numerical tool to address the problem of the
electron-electron (e-e) interactions is the
time-dependent-density-matrix-renormalization-group\cite{tddmrg}\cite{tddmrg2}\cite{tddmrg3},
which turns out to have excellent performance in one-dimensional
(1D) systems. For what concerns electronic systems with
time-dependent parameters but not involving charge transport, in a
recent paper Cazalilla carried out the exact evolution of a
noninteracting 1D system following a sudden switch-on of Luttinger
liquid e-e interaction\cite{caz}. Remarkably, it is found that the
initial ground state reaches a stationary state only if the system
is infinite-sized and that its asymptotic behavior differs from
the one of the interacting equilibrium system\cite{rigol}.

In the present paper we extend this analysis to two coupled
interacting 1D systems. Our
system consists of two isolated Luttinger liquids which are
connected at $t=0$ through interactions between chiral electron
densities located at different liquids.

The Luttinger liquid is the prototype of interacting electrons in 1D
and it is governed by the so called Tomonaga-Luttinger Hamiltonian.
In this model, the electrons have linear dispersion relation around
positive (Right) and negative (Left) Fermi points located at $\pm k_{F}$
and the e-e interactions act only
between Right/Left electron densities. This means that only the component
with momentum transfer $p \sim 0$ of the Coulomb repulsion is
retained, while the $p \sim 2k_F$ component is assumed to be
negligible. The model is exactly solvable by means of bosonization
technique, which allows to write the electron Hamiltonian in terms
of boson operators $b$'s.

%%%%%%%%%%%%%%%%%%%%%%%%%%%%%%%%%%%%%%%%%%%%%%%%%%%%%%%%%%%%%%%%%%%
%%%%%%%%%%%%%%%%%%%%%%%%%%%%%%%%%%%%%%%%%%%%%%%%%%%%%%%%%%%%%%%%%%%

In the following we shall adopt a similar notation as in Ref.
\onlinecite{caz}.

For $t<0$ the system consists of two disconnected identical
interacting electron liquids described by the bosonized
Hamiltonian\cite{gdsv} $H_0=H_1+H_2$ with
\begin{eqnarray}
H_{i}=\frac{1}{2}\sum_{q \neq 0}&&[(v_F
+g^{(4)}_4(q))|q|(b^{\dagger}_i(q) b_i(q)+b_i(q) b^{\dagger}_i(q))
\nonumber \\&& - g^{(2)}_4(q)
(b^{\dagger}_i(q)b^{\dagger}_i(-q)+b_i(q)b_i(-q)) ]
\end{eqnarray}
where $i=1,2$,
$[b_i(q),b^{\dagger}_j(q')]=\delta_{i,j}\delta_{q,q'}$, $v_F$ is the
Fermi velocity, $g_{4}^{(4)}(q)$ is the interaction parameter
between  Right-Right (positive $q$) and Left-Left (negative $q$)
electron densities, while $g_{4}^{(2)}(q)$ is the interaction
parameter between Left-Right densities.
%It is worth to recall that the boson
%operators $b$'s are related to the original electron operators $c$'s
%by the following expression:
%\begin{eqnarray}
%b_{i}(q)=\sqrt{\frac{2 \pi}{L |q|}}\sum_{k \neq
%0}c^{\dagger}_{i}(k+q)c_{i}(k) \nonumber \\
%b_{i}(-q)=\sqrt{\frac{2 \pi}{L |q|}}\sum_{k \neq
%0}c^{\dagger}_{i}(k)c_{i}(k+q)
%\end{eqnarray}

The system is diagonalized by the well-known Bogolioubov
transformation:
\begin{eqnarray}
\tilde{b}_i(q)&=&\cosh \varphi (q) b_i(q)+\sinh \varphi (q)
b^{\dagger}_i(-q) \, , \nonumber \\
\tilde{b}^{\dagger}_i(q)&=&\sinh \varphi (q) b_i(-q)+\cosh \varphi
(q) b^{\dagger}_i(q) \, \label{bog1}
\end{eqnarray}
with
\begin{eqnarray}
\tanh 2 \varphi(q) = g_4^{(2)}(q)/[v_F+g_4^{(4)}(q)] \,
,\label{tan1}
\end{eqnarray}
%The diagonal form of $H_i$ the reads
%\begin{eqnarray}
%H_{i}=\frac{1}{2}\sum_{q \neq 0}v(q)|q|\tilde{b}^{\dagger}_i(q)
%\tilde{b}_i(q)  \,
%\end{eqnarray}
and renormalized velocity
\begin{equation}
v(q)=\sqrt{(v_F +g^{(4)}_4(q))^2-(g_4^{(2)}(q))^2 } \, .
\end{equation}

For $t \geq 0$ the  chiral density-density interactions between the
two Luttinger liquids
 are switched on and the system is governed by the total Hamiltonian
\begin{eqnarray} H_{tot}=H_0+\theta(t)H_{12}
\end{eqnarray}
where
\begin{eqnarray}
H_{12}=&&\sum_{q \neq 0}[g^{(4)}_2(q)(b^{\dagger}_1(q) b_2(q)+
b^{\dagger}_2(q) b_1(q) \nonumber \\
+&& b_1(q) b^{\dagger}_2(q) +b_2(q) b^{\dagger}_1(q)
  \nonumber \\
- &&g^{(2)}_2(q) (b^{\dagger}_1(q)b^{\dagger}_2(-q)+b_1(q)b_2(-q)
\nonumber \\
+ &&b^{\dagger}_2(q)b^{\dagger}_1(-q)+b_2(q)b_1(-q))] \, ,
\end{eqnarray}
where $g_{2}^{(4)}(q)$ ( $g_{2}^{(2)}(q)$ ) is the interaction
parameter between electron densities of the same (opposite)
chirality  in different liquids. Physically $H_{12}$ could
represent the long-range component of the Coulomb repulsion felt
by electrons located in two (quasi)1D metallic systems close to
each other. Here we are considering a general situation where
$g^{(4)}_2(q) \neq g^{(2)}_2(q)$ but in real systems like carbon
nanotubes, they coincide (see below). It is worth to remark that
in the present model the coupling between the two liquids does not
involve any inter-liquid electron tunneling.

$H_{tot}$ is again diagonalized by introducing symmetrized and
antisymmetrized operators\cite{komnik}:
\begin{eqnarray}
s(q)&=&[b_1(q)+b_2(q)]/\sqrt{2} \, , \nonumber  \\
a(q)&=&[b_1(q)-b_2(q)]/\sqrt{2} \, .
\end{eqnarray}
In terms of them $H_{tot}$ decouples in two independent
non-equivalent Luttinger liquids:
\begin{eqnarray}
&&H_{tot}=H_{s}+H_{a} \nonumber \\
&& = \frac{1}{2}\sum_{q \neq
0}[(v_{F}+g^{(4)}_4(q)+g^{(4)}_2)|q|(s^{\dagger}(q)
s(q)+s(q) s^{\dagger}(q))  \nonumber \\
&& - (g^{(2)}_4(q)+g^{(2)}_2(q)) (s^{\dagger}(q)s^{\dagger}(-q)+s(q)s(-q))] \nonumber \\
&& + \frac{1}{2}\sum_{q \neq
0}[(v_{F}+g^{(4)}_4(q)-g^{(4)}_2)|q|(a^{\dagger}(q)
a(q)+a(q) a^{\dagger}(q)) \nonumber\\
&& -
(g^{(2)}_4(q)-g^{(2)}_2(q))(a^{\dagger}(q)a^{\dagger}(-q)+a(q)a(-q))
] \, .
\end{eqnarray}
Finally, the diagonal form of $H_{tot}$ is obtained in terms of
\begin{eqnarray}
\tilde{s}(q)&=&\cosh \varphi_{s} (q) s(q)+\sinh \varphi_{s} (q)
s^{\dagger}(-q) \, , \nonumber \\
\tilde{s}^{\dagger}(q)&=&\sinh \varphi_{s} (q) s(-q)+\cosh
\varphi_{s} (q) s^{\dagger}(q) \,  , \nonumber \\
\tilde{a}(q)&=&\cosh \varphi_{a} (q) a(q)+\sinh \varphi_{a} (q)
a^{\dagger}(-q) \, ,  \nonumber \\
\tilde{a}^{\dagger}(q)&=&\sinh \varphi_{a} (q) a(-q)+\cosh
\varphi_{a} (q) a^{\dagger}(q) \, , \label{bog2}
\end{eqnarray}
where
 \begin{equation} \tanh 2 \varphi_{s,a}(q) = [g_4^{(2)}(q) \pm
g_2^{(2)}(q) ]/[v_F+g_4^{(4)}(q) \pm g_2^{(4)}(q)] \, , \label{tan2}
\end{equation}
%Therefore it holds
%\begin{eqnarray}
%H_{tot}=&&\frac{1}{2}\sum_{q \neq
%0}v_{s}(q)|q|\tilde{s}^{\dagger}(q) \tilde{s}(q) + \nonumber \\
%&&\frac{1}{2}\sum_{q \neq 0}v_{a}(q)|q|\tilde{a}^{\dagger}(q)
%\tilde{a}(q)
% \,
%\end{eqnarray}
and renormalized velocities
\begin{equation}
v_{s,a}(q)=\sqrt{(v_F+g_4^{(4)}(q) \pm g_2^{(4)}(q))^2-(g_2^{(2)}(q)
\pm g_4^{(2)}(q))^2 }\, .
\end{equation}

Now we are in the position to evaluate the equal-time one-particle
correlation function for $t>0$, defined as
\begin{eqnarray}
G^{(i)}_{\gamma}(x,t)=\langle e^{i H_{tot}t} \psi_{i,\gamma}(x)
\psi^{\dagger}_{i,\gamma}(0) e^{-i H_{tot}t} \rangle _{H_{0}} \,
\end{eqnarray}
where $i=1,2$ labels the two  Luttinger liquids decoupled at $t<0$,
the subscript $\gamma=R,L$ indicates the Right/Left character of the
non-interacting fermion fields $\psi$ and $\psi ^{\dagger}$, and
$\langle \ldots \rangle _{H_{0}}$ is the zero-temperature average in
the ensemble $H_{0}$. Without any loss of generality we shall focus
on $G^{(1)}_{R}(x,t)$.

The bosonization technique allows to calculate $G_{R}$ with
logarithmic accurancy, which is exact for distances/times much
longer than the typical range of the interactions. The key point of
bosonization is that it is possible to express the fermion field in
terms of boson fields. For instance the Right-mover fermion field is
given by the following expression:
\begin{eqnarray}
\psi_{i,R}(x)=\frac{\eta_{R}}{(2 \pi \alpha)^{1/2}}e^{i
\Phi_{i,R}(x)} \, ,
\end{eqnarray}
where $\alpha$ is a short-distance cutoff, proportional to the
lattice spacing, $\eta_{R}$ is an anticommuting Klein factor and
\begin{eqnarray}
\Phi_{i,R}(x) &=& \sum_{q>0} \left( \frac{2\pi}{qL} \right)^{1/2}
e^{-\alpha q/2} [b_{i}^{\dagger}(q) e^{-iqx}+ b_{i}(q) e^{iqx}]
\nonumber \\  &+& \varphi_{0,R}+ 2\pi x  N_R /L  \, ,
\end{eqnarray}
where $N_R$ is the total number of Right-electrons, $[\varphi_{0,R},
N_R ]=i$ and $L$ is the length of the system.

 Thus the computation of the correlation function reduces to:
\begin{eqnarray}
G^{(1)}_{R}(x,t)=\frac{1}{2 \pi \alpha} \langle e^{i
\Phi_{1,R}(x,t)} e^{-i \Phi_{1,R}(0,t)} \rangle _{H_{0}} = \nonumber
\\  \frac{1}{2 \pi \alpha}
 e^{-\frac{1}{2} \left\{ \langle (\Phi_{1,R}(x,t)-  \Phi_{1,R}(x,t))^{2} \rangle
_{H_{0}} - [\Phi_{1,R}(x,t),\Phi_{1,R}(0,t)] \right\} }
\end{eqnarray}
where $\Phi_{1,R}(x,t)= e^{i H_{tot}t} \Phi_{1,R}(x) e^{-i
H_{tot}t}$. In order to compute $\Phi_{1,R}(x,t)$ it is convenient
to evaluate first  $b_1(q,t)= e^{i H_{tot}t} b_1(q) e^{-i H_{tot}t}$
in terms of the $\tilde{b}_i(q)$'s and $\tilde{b}^{\dagger}_i(q)$'s,
since they diagonalize $H_0$ which defines the ensemble we average
on.

After a lengthy algebra involving a direct and an inverse
Bogolioubov transformation in Eq. \ref{bog2} and the inverse of Eq.
\ref{bog1}, one finds
\begin{eqnarray}
b_1(q,t)&=&A(q,t)\tilde{b}_1(q)+B^{\star}(q,t)\tilde{b}^{\dagger}_1(-q)
\nonumber \\
&+&C(q,t)\tilde{b}_2(q)+D^{\star}(q,t)\tilde{b}^{\dagger}_2(-q) \, ,
\end{eqnarray}
where
\begin{eqnarray}
A(q,t)&=&\cosh \varphi(q) \left[- i \sin v_s |q| t \cosh
2\varphi_s(q) \right.
\nonumber \\
&& \left. + \cos v_s |q| t + (a \leftrightarrow s) \right] \nonumber \\
&& -\sinh \varphi(q) \left[- i \sin v_s |q| t \sinh 2\varphi_s(q)
\right.
\nonumber \\
&& \left.  + (a \leftrightarrow s) \right] \, , \nonumber \\
B^{\star}(q,t)&=&-\sinh \varphi(q) \left[- i \sin v_s |q| t \cosh
2\varphi_s(q) \right.
\nonumber \\
&& \left. + \cos v_s |q| t + (a \leftrightarrow s) \right] \nonumber \\
&& +\cosh \varphi(q) \left[- i \sin v_s |q| t \sinh 2\varphi_s(q)
\right.
\nonumber \\
&& \left.  + (a \leftrightarrow s) \right] \, , \nonumber \\
C(q,t)&=&\cosh \varphi(q) \left[- i \sin v_s |q| t \cosh
2\varphi_s(q) \right.
\nonumber \\
&& \left. + \cos v_s |q| t - (a \leftrightarrow s) \right] \nonumber \\
&&-\sinh \varphi(q) \left[- i \sin v_s |q| t \sinh 2\varphi_s(q)
\right.
\nonumber \\
&& \left.  - (a \leftrightarrow s) \right] \, , \nonumber \\
D^{\star}(q,t)&=&-\sinh \varphi(q) \left[- i \sin v_s |q| t \cosh
2\varphi_s(q) \right.
\nonumber \\
&& \left.  + \cos v_s |q| t - (a \leftrightarrow s) \right] \nonumber \\
&& +\cosh \varphi(q) \left[- i \sin v_s |q| t \sinh 2\varphi_s(q)
\right.
\nonumber \\
&& \left. - (a \leftrightarrow s) \right] \, .
\end{eqnarray}
The above relations allow to evaluate exactly the time-dependence in
the correlation function, which reads:
\begin{eqnarray}
&& G^{(1)}_{R}(x,t)=  \frac{1}{2 \pi \alpha}
 \exp [ -\sum_{q>0} \left( \frac{2\pi}{qL} \right)
e^{-\alpha q}  \nonumber \\
&& \times  [(|A(q,t)|^{2}+|C(q,t)|^{2})(-i\sin qx+1-\cos qx)
\nonumber \\
&&  + (|B(q,t)|^{2}+|D(q,t)|^{2})(i\sin qx+1-\cos qx)]]
\end{eqnarray}
The sum over $q$ is performed by using that $q=2 \pi n/L $ and we
end up with
\begin{eqnarray} &&
G^{(1)}_{R}(x,t)= \frac{c}{d(x) ^{\frac{1}{4}(2 \cosh 2\varphi+
\cosh (2\varphi-4\varphi_s) +  \cosh (2\varphi-4\varphi_a))}} \nonumber \\
&& \times  \left[ \frac{d(x-2v_{s}t)d(x+2v_{s}t)}{d(2v_{s}t)^{2}}
\right] ^{\frac{1}{8}(\cosh
(2\varphi-4\varphi_s) - \cosh 2\varphi  )} \nonumber \\
&& \times  \left[ \frac{d(x-2v_{a}t)d(x+2v_{a}t)}{d(2v_{a}t)^{2}}
\right] ^{\frac{1}{8}( \cosh (2\varphi-4\varphi_a)- \cosh 2\varphi)}
\label{gfin}
\end{eqnarray}
where $d(x)=L \sin  (\pi x/L)  $. In order to obtain the above
expression we have introduced momentum-independent $\varphi$ and
$\varphi_{s,a}$ such that $\sinh f (q) \approx e^{-R_0 |q|/2} \sinh
f (0)$ with $R_0$ of the order of the range of the e-e
interactions; in this way  the finite overall prefactor $c$ is
produced\cite{lut}.

Let us now consider some relevant limits.

First, it is straightforward to check that in the non-interacting
case (\textit{i.e.} $\varphi=\varphi_{s}=\varphi_{a}=0$ and $v_a =
v_s = v_F$)  we recover
\begin{eqnarray} &&
G^{(1)}_{R}(x,t)= \frac{c}{d(x)} \, ;
\end{eqnarray}
on the other hand for $t \rightarrow 0$ we find
\begin{eqnarray} &&
G^{(1)}_{R}(x,0)= \frac{c'}{d(x)^{\cosh 2 \varphi}} \, ,
\label{disc}
\end{eqnarray}
where $c'$ is again a finite constant; this result corresponds to
the correlation function of an individual Luttinger liquid governed
by $H_1$ or $H_2$, as it should.

More interesting is the case in which $g_4^{(2)}=g_4^{(4)}=g_2^{(4)}=0$
(\textit{i.e.} $\varphi=0$ and $-\varphi_{a}=\varphi_{s} \equiv
\bar{\varphi} \neq 0  $ and $v_a = v_s \equiv \bar{v} \neq v_F$). In
this case the system corresponds to have we two identical replicas
of the model considered by Cazalilla, \textit{i.e.} an isolated noninteracting
1D system with density-density interactions switched on at $t=0$. In
this case the we obtain
\begin{eqnarray} &&
G^{(1)}_{R}(x,t)= \frac{c}{d(x) ^{\frac{1}{4}(
\cosh 4\bar{ \varphi} +1) }} \nonumber \\
&& \times  \left[ \frac{d(x-2 \bar{v} t)d(x+2 \bar{v} t)}{d(2
\bar{v} t)^{2}} \right] ^{\frac{1}{4}( \cosh 4 \bar{\varphi}-1) }
\end{eqnarray}
which recovers the result of Cazalilla (Eq. 9 of Ref.
\onlinecite{caz}).

Anyway the most remarkable result is obtained from the long-time
limit $t \rightarrow\infty$ of Eq. \ref{gfin}. In this case we get
\begin{equation}
G^{(1)}_{R}(x) \approx \frac{c}{|x| ^{\frac{1}{4}(2 \cosh 2\varphi+
\cosh (2\varphi-4\varphi_s) +  \cosh (2\varphi-4\varphi_a))}}
\label{asi}
\end{equation}
where we have also taken $L \rightarrow \infty$. We point out that
if the infinite length limit is not taken, the correlations obey
an oscillatory behavior even at long time, confirming that in a
finite-sized system a steady state is not reachable. However, as
already discussed in Ref. \onlinecite{caz}, in mesoscopic systems
with finite $L$ (e.g. $L \sim 1 \mu$m for carbon nanotubes) a
genuine stationary state can be reached because of finite
temperature effects. Indeed if the temperature $T$ is larger than
the level spacing $\Delta \varepsilon \sim v_{F}/L$, we have to
replace $L$ by $v_{F}/T$ and $\sin$ by $\sinh$ in the definition
of $d(x)$. As a consequence for very large distances/times the
Green function decays exponentially as $G_{R}^{1} \sim e^{-|x|\nu
T}$, where $\nu$ is the same exponent displayed in the power-law
at $T=0$\cite{giamarchi}. Thus it appears that in the exponential
regime we can still keep track of the Luttinger liquid
correlations encoded by $\nu$. On the other hand for
distances/times smaller than the thermal length $v'_{F}/T$ (with
$v'_{F}$ the renormalized Fermi velocity\cite{giamarchi}) the
power-law discussed above still holds.

In the thermodynamic limit, it is worth to notice that the  result
in Eq. \ref{asi} does not coincide with the equilibrium equal-time
correlation function of a two coupled Luttinger liquids governed
by $H_0+H_{12}$. Indeed in the latter case the correlation
function reads
\begin{equation}
G^{(1)}_{R}(x) \approx \frac{c''}{|x| ^{\frac{1}{2}( \cosh
(2\varphi_s) +  \cosh (2\varphi_a))}} \, , \label{equi}
\end{equation}
with finite $c''$. The  difference in the critical exponents in the
two cases reveals that in the long-time limit the ground state of
$H_0$ evolved by $H_0+\theta (t) H_{12}$ reaches a stationary state
which is \textit{not} the ground state of $H_0+H_{12}$. This is due
to the fact that for $t>0$ the energy is conserved and therefore the
system cannot relax to the ground state. Indeed the critical
exponent in Eq. \ref{asi} is larger than the critical exponent in
Eq.  \ref{equi}, consistently with the fact that in the latter case
the system is able to optimize the repulsive interactions given by
$H_{12}$.

Now let us discuss physical consequences of this result. To this end
we introduce the critical exponent $\alpha$ governing some relevant
observables like the tunneling density of states $\rho$\cite{voit1}
\begin{equation}
\rho (\omega) \sim  \omega^{\alpha}
\end{equation}
which is detectable  with scanning tunneling microscopy by probing
the bulk the 1D conductor\cite{eg}\cite{kane}.

In order to compare to real systems, we have to introduce spin in
the previous analysis. This is completely strightforward as long as
we consider spin-independent e-e interactions. For
spinless electrons, the $\alpha$ exponent is obtained by subtracting
1 to the exponent one-particle correlation function, while for the
corresponding spinful system we just have to half such a
value\cite{voit}.

Anyway it is worth to note that, while in equilibrium systems
the critical exponent of $\rho(\omega)$
can be safely extracted from the
equal-time one-particle Green function, in out-of-equilibrium
situations such a procedure is far from obvious\cite{kg}.
%since
%$G^{i}_{\gamma}(x,t,t') \neq G^{i}_{\gamma}(x,t-t')$.
Therefore it would be convenient to relate $\alpha$ to some other
energy-independent observable, like the momentum distribution,
which also displays a power law\cite{voit1}:
\begin{equation}
n(q) \propto \mathrm{sign} (q-k_{F}) |q-k_{F}|^{\alpha} \, .
\end{equation}
$n(q)$ is accessible, for instance, from
angle-resolved-photo-emission-spectroscpy, which has been
successfully applied to quasi-1D
meterials\cite{arpes3}\cite{arpes1} to measure
$\alpha$\cite{arpes2}\cite{arpes4}. We also mention that such a
technique is not affected by the uncertanties regarding the
contacts, which are present in transport experiments. At this
point, we note again that the above power-law  would be cutoff by
finite temperature effects if $|q-k_{F}| \sim T/v_{F}$.

From Eqs. \ref{asi} and \ref{equi} the exponent governing $n(q)$
in the two different cases  discussed above read
\begin{eqnarray}
&& \alpha_{\mathrm{asympt}}= \nonumber \\
&& \frac{1}{8}(2 \cosh 2\varphi+ \cosh (2\varphi-4\varphi_s) +
\cosh (2\varphi-4\varphi_a))-\frac{1}{2} \, , \nonumber \\
&& \alpha_{\mathrm{equil}}= \frac{1}{4}( \cosh (2\varphi_s) + \cosh
(2\varphi_a))-\frac{1}{2} \, . \label{esponenti}
\end{eqnarray}

The model we have considered could find an experimental
realization in 1D fermionic systems built within optical
potential. Unfortunately in these systems the interaction is
short-ranged and this makes the term $H_{12}$ hard to realize.
Anyway long-range dipole-dipole interactions have been recently
obtained in chromium bosonic atoms\cite{longrange} and Fermi gases
with long-range interactions are likely becoming available.

On the other hand it is tempting to relate our results to the
physics of carbon nanotubes. Metallic carbon nanotubes are
believed to be rather good (although approximate) realizations of
Luttinger liquids since in normal conditions the main correlation
effects come from the long-range part of the Coulomb repulsion
(through the $g_{i}^{(j)}$ parametres), while the back-scattering
interactions with large momentum transfer suffer a $1/R$
suppression, where $R$ is the radius of the
nanotube\cite{eg}\cite{kane}. Indeed back-scattering interactions
are marginal operators in the renormalization group sense and in
carbon nanotubes are predicted to breakdown the Luttinger liquid
state only at very low energy scales\cite{eg}. This is supported
by direct observation of power-law decay of the linear conductance
as a function of temperature\cite{gao}\cite{yao}\cite{bock}.

It is worth to recall that each individual carbon nanotube is composed
itself by two coupled (identical) Luttinger liquids, since there are
a Left and a Right branch respectively around each of the two Fermi
points\cite{komnik}\cite{eg}\cite{gonzperf}. This implies that the
critical exponent governing the decay of the correlation functions
in these systems is actually one half the exponents defined in Eqs.
\ref{esponenti}. Therefore in the following we will use
$\alpha_{\mathrm{asympt}}^{\mathrm{(nano)}}=\alpha_{\mathrm{asympt}}/2$
and
$\alpha_{\mathrm{equil}}^{\mathrm{(nano)}}=\alpha_{\mathrm{equil}}/2$.

The idea is to compare two possible measurements realized in two different
experimental setups. The first system (setup A), which gives access
to $\alpha_{\mathrm{asympt}}^{\mathrm{(nano)}}$, is made by two
(initially distant) nanotubes which can be quickly aligned at controlled
distance. This makes possible to mimic
the switching on of inter-tube interactions.
In this case, of course, the critical exponent could be only be
measured in the final steady state and
any transient regime is definitively unaccessible.
The second system
(setup B), which gives access to
$\alpha_{\mathrm{equil}}^{\mathrm{(nano)}}$, consists in having the
carbon nanotubes synthetized at a given distance, for instance in a
double-wall nanotube.
In this case the system has fixed inter-tube interactions
built-in from the very beginning.

Typical metallic nanotubes correspond to armchair $(10,10)$ geometry
(\textit{i.e.} radius $R \approx 7$\AA $\,$) with length $L \sim  1
\mu$m. The strength of the density-density Coulomb interactions is
then\cite{eg}\cite{gonzperf}
\begin{eqnarray}
g^{(2)}_{4}(q \sim 0)=g^{(4)}_{4}(q \sim 0) \approx \frac{4
e^{2}}{\kappa \pi v_F} K_0 (q_{c}R ) \, ,\nonumber \\
g^{(2)}_{2}(q \sim 0)=g^{(4)}_{2}(q \sim 0) \approx \frac{4
e^{2}}{\kappa \pi v_F} K_0 (q_{c}D)  \, ,\nonumber \\
\label{gpar}
\end{eqnarray}
where $K_0$ is the modified Bessel function, $q_c = 2\pi/L$ is the
infrared momentum cutoff due to the finite size of the nanotubes,
$D \gtrsim R$ is the intertube distance, $\kappa \sim 2$ is the
dielectric constant of typical nanotubes and $v_F \approx 8 \times
10^{5}$m/s. We recall at this point that the amplitude of
intra-tube backscattering interactions is approximatively\cite{eg}
$0.1 e^{2} a/ \kappa \pi v_{F} R$ (with $a = 2.46 $\AA) which is
much smaller of both intra- and inter-tube forward scattering
interactions for $D\gtrsim R$. We also mention that
single-electron tunneling between nanotubes is usually strongly
suppressed since the misalignment of the respective carbon
lattices prevents momentum conservation in the tunneling process.
Therfore the experimental setups that we propose could be fairly
approximated  by the model of two coupled Luttinger liquids.

By inserting $g^{(j)}_{i}$ obtained in Eqs. \ref{gpar} in Eqs. \ref{tan1} and
\ref{tan2}, we can estimate the critical exponents accessible to
experiments within setups A and B.

\begin{figure}
\begin{center}
\mbox{\epsfxsize 8.5cm \epsfbox{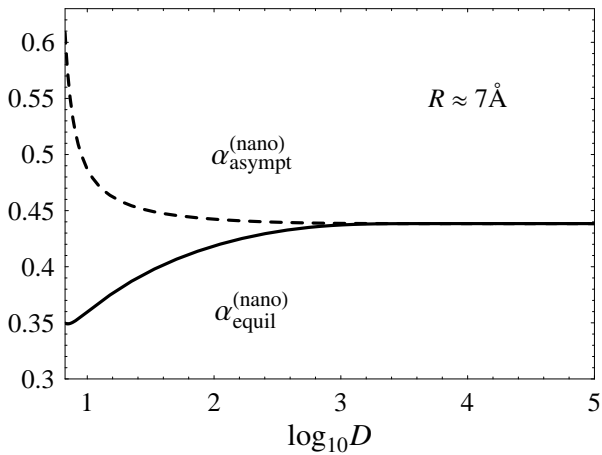}}
\end{center}
\caption{Plot of the critical exponents
$\alpha_{\mathrm{asympt}}^{\mathrm{(nano)}}$ (dashed curve) and
$\alpha_{\mathrm{equil}}^{\mathrm{(nano)}}$ (solid curve) for
typical $(10,10)$ nanotubes as a function of the logarithm of the
distance $D$ between the nanotubes. In $\log _{10}D$, $D$ is
expressed in \AA.} \label{figure}
\end{figure}

The two exponents are plotted in Fig. \ref{figure} as a function of
the distance between the nanotubes. For $D \rightarrow \infty$ the
intertube couplings $g^{(4)}_{2}$ and $g^{(2)}_{2}$ vanish (see Eq.
\ref{gpar}) and $\alpha_{\mathrm{asympt}}^{\mathrm{(nano)}}$ and
$\alpha_{\mathrm{equil}}^{\mathrm{(nano)}}$ tend to the common value
\begin{eqnarray}
&& \alpha_{\mathrm{disc}}^{\mathrm{(nano)}}=\frac{1}{4} (\cosh 2
\varphi -1 ) \approx 0.43 \, ,
\end{eqnarray}
which governs the correlations of two completely independent
nanotubes, according to Eq. \ref{disc}. This value is in agreement
with previous estimates given for single typical
nanotubes\cite{eg}\cite{kane}.

When the two nanotubes are brought close to each other in setup A
(Fig. 1 dashed line), the intertube interaction are switched on
and
$\alpha_{\mathrm{asympt}}^{\mathrm{(nano)}}$ increases. In particular at  %$D \sim 10$\AA $\,$
$D \sim R$ we find
\begin{eqnarray}
&& \alpha_{\mathrm{asympt}}^{\mathrm{(nano)}} \approx 0.60 \, .
\end{eqnarray}
Regarding setup B (Fig. 1 solid line), if the two carbon nanotubes
have been  synthetized at the same distance %$D \sim 10$\AA $\,$,
$D \sim R$, the predicted
exponent is
\begin{eqnarray}
&& \alpha_{\mathrm{equil}}^{\mathrm{(nano)}} \approx 0.35 \, ,
\end{eqnarray}
that is significantly smaller than
$\alpha_{\mathrm{asympt}}^{\mathrm{(nano)}}$. We believe that such
a large difference could be detected, even though the appropriate
experimental setup might be difficult to realize. In particular in
setup A the two nanotubes should approach each other at a speed
comparable to (or even higher than) the Fermi velocity ($\sim
10^{6}$m/s). We note, by the way, that this problem does not arise
in cold atomic systems, where the Fermi velocity is orders of
magnitude smaller.

%It is tempting to relate these estimates to the measurements of the
%critical exponent carried out in two experiments resembling to the
%ones that we propose. In Ref. \onlinecite{gao} the critical exponent
%of two crossed carbon nanotubes  with negligible intertube tunneling
%was measured. This instance may reproduce approximatively the setup
%A and the measurement of the tunneling conductance gives
%$\alpha_{\mathrm{crossed}} \approx 0.55$\cite{note}. Anyway we must
%recall that in the present analysis we are disregarding some
%possibly relevant feature of the setup of Ref. \onlinecite{gao},
%like the enhancement of backscattering\cite{komnik} due to the
%mechanical deformation of the tubes at the crossing region.
%
%On the other hand in Ref. \onlinecite{bock} it was measured the
%exponent of a bulk-contacted rope of nanotubes characterized by
%$\alpha_{\mathrm{rope}} \approx 0.33- 0.38$, which can be
%assimilated to setup B.

In conclusion we have computed the time-dependent evolution of the
single-particle correlation function of two Luttinger liquids
coupled by a sudden switch-on of the inter liquid interaction. This
allows to evaluate the critical exponent $\alpha$ governing some
physical observables accessible to real experiments, such as the
momentum distribution function. We find
that the in the long-time limit the initial ground state relaxes to
a stationary state which is not the equilibrium ground state of the
coupled system. In the latter case the critical exponent
$\alpha_{\mathrm{equil}}$ results to be smaller than the asymptotic
exponent $\alpha_{\mathrm{asympt}}$ of about a factor 2.
An experiment capable to detect such a remarkable finding in carbon nanotubes is proposed.
Finally we believe that the present study could be also
relevant in experiments involving ultracold fermionic atoms loaded
in optical lattices, where tunable Luttinger liquids
might be realized in the near future.

The author kindly acknowledges helpful discussions with J.
Gonz{\'a}lez, M. Cini and G. Stefanucci. This work has been supported
by Ministerio de Educaci\'on y Ciencia (Spain) through grants
FIS2005-05478-C02-01/02 and by INFN (Italy) through grant 10068.


\begin{thebibliography}{99}

\bibitem{prog}
T. Fujisawa, T. Hayashi and S. Sasaki, Rep. Prog. Phys. \textbf{69},
759 (2006).

\bibitem{timeqd}
T. Fujisawa, Y. Tokura and Y. Hirayama, Phys. Rev. B \textbf{63},
R081304 (2001).

\bibitem{opt1}
M. Greiner \textit{et al.}, Nature (London) \textbf{415}, 39 (2002); T.
St\"{o}ferle \textit{et al.}, Phys. Rev. Lett. \textbf{92}, 130403 (2004).

\bibitem{opt2}
H. Ott \textit{et al.}, Phys. Rev. Lett. \textbf{92}, 160601 (2004);
M. K\"{o}hl \textit{et al.}, Phys. Rev. Lett. \textbf{94}, 080403
(2005).

\bibitem{hcb}
T. Kinoshita, T. Wenger, and D. S. Weiss,  Nature (London)  {\bf
440}, 900 (2006).

\bibitem{rigol}
M. Rigol, V. Dunjko, V. Yurovskii, and M. Olshanii, cond-mat/0604476
(2006).

\bibitem{cini} M. Cini, Phys. Rev. B {\bf
22}, 5887 (1980).

\bibitem{stef} G. Stefanucci and C. O. Almbladh,
Phys. Rev. B {\bf 69}, 195318
 (2004).

\bibitem{tddmrg}
M. A. Cazalilla and J. B. Morton, Phys. Rev. Lett. {\bf 88}, 256403
(2002).

\bibitem{tddmrg2}
S. R. White, and A. E. Feiguin, Phys. Rev. Lett. {\bf 93}, 076401
(2004).

\bibitem{tddmrg3}
A.J. Daley,  C. Kollath, U. Schollw\"{o}ck and G. Vidal,
J. Stat. Mech.: Theor. Exp. P04005 (2004)

\bibitem{caz}
M. A. Cazalilla, cond-mat/0606236 (2006).

\bibitem{gdsv}
J. Gonz\`{a}lez \textit{et al.}, \textit{Quantum Electron Liquids
and High-$T_c$ Superconductivity}, Chap. 4, Springer-Verlag, Berlin
(1995).

\bibitem{komnik}
A. Komnik, and R. Egger, Phys. Rev. Lett. {\bf 80} 2881 (1998).

\bibitem{lut}
A. Luther and I. Peschel, Phys. Rev. B {\bf 9} 2911 (1974).

\bibitem{giamarchi}
T. Giamarchi, \textit{Quantum Physics in One Dimension}, Chap. 3, Oxford
Science Publications, Oxford
(2004).

\bibitem{voit1}
J. Voit, J. Phys.: Cond. Matt.  {\bf 5}, 8305  (1993).

%\bibitem{kanefish}
%C. Kane, and M. Fisher, Phys. Rev. B {\bf 68}, 1220 (1992).

\bibitem{eg}
R. Egger and A. O. Gogolin, Phys. Rev. Lett. {\bf 79}, 5082 (1997);
Eur. Phys. J. B {\bf 3}, 281 (1998).

\bibitem{kane}
C. Kane, L. Balents, and M. Fisher, Phys. Rev. Lett. {\bf 79}, 5086
(1997).

\bibitem{kg} A. Komnik and A. O. Gogolin, Phys. Rev. B {\bf 66}, 125106
(2002).

\bibitem{arpes3}
J. Voit \textit{et al.}, Science  {\bf 290}, 501, (2000).

\bibitem{arpes1} J. Choi, S.M. Lee, Y.C. Choi, Y.H. Lee, J.C. Jiang,
 Chem. Phys. Lett., {\bf 349}, 185 (2001).

\bibitem{arpes2}
H. Ishii \textit{et al.}, Nature
(London)  {\bf 426}, 540, (2003).

\bibitem{arpes4} We note that in Ref. \onlinecite{arpes2} the photoemission
experiment is not angle-resolved.

\bibitem{voit}
J. Voit, Rep. Prog. Phys. {\bf 58}, 977 (1995).

\bibitem{longrange}J. Stuhler, A. Griesmaier, T. Koch, M. Fattori, T. Pfau, S. Giovanazzi, P. Pedri, and L. Santos,
Phys. Rev. Lett. \textbf{95}, 150406 (2005).

\bibitem{gao}
B. Gao, A. Komnik, R. Egger, D. C. Glattli, and A. Bachtold, Phys.
Rev. Lett. {\bf 92} 216804 (2004).

\bibitem{yao}
Z. Yao, H. W. Postma, L. Balents, and C. Dekker, Nature (London)
{\bf 402}, 273 (1999).

\bibitem{bock}
M. Bockrath, D. H. Cobden, J. Lu, A. G. Rinzler, R. E. Smalley, L.
Balents and P. L. McEuen,   Nature {\bf 397}, 598 (1999).

\bibitem{gonzperf}
J. Gonz\`{a}lez and E. Perfetto, Phys. Rev. B {\bf 72}, 205406
(2005);  Eur. Phys. J.B \textbf{51}, 571 (2006).

%\bibitem{delaney}
%P. Delaney, H. J. Choi, J. Ihm, S. G. Louie, and M. L. Cohen, Nature
%(London)  {\bf 391}, 466, (1998).

%\bibitem{liu}
%J. Liu, M.l J. Casavant, M. Cox, D. A. Walters, P. Boul, W. Lu, A.
%J. Rimberg, K. A. Smith, Daniel T. Colbert, and Richard E. Smalley,
%Chem. Phys. Lett. {\bf 303}, 125 (1999).

%\bibitem{note} In Ref. \onlinecite{gao} it was measured the
%bulk-bulk tunneling exponent, which is
%$\alpha_{\mathrm{bulk-bulk}}\approx 1.1$. Anyway such a  value is
%the double of the bulk exponent which is consider in the present
%paper\cite{eg}\cite{kane}.










\end{thebibliography}
\end{document}